\documentclass[%
 reprint,
superscriptaddress,
nofootinbib,
 amsmath,amssymb,
aps,
pra,
]{revtex4-1}

\usepackage{graphicx}% Include figure files
\usepackage{dcolumn}% Align table columns on decimal point
\usepackage{bm}% bold math
\usepackage{hyperref}% add hypertext capabilities
\begin{document}

\preprint{APS/123-QED}

\title{Quantum memory assisted precision rotation sensing}% Force line breaks with \\

\author{Wenkui Ding}
\email{wenkuiding@zju.edu.cn}
\affiliation{Zhejiang Institute of Modern Physics and Department of Physics,
Zhejiang University, Hangzhou, Zhejiang 310027, China}

\author{Wenxian Zhang}
\affiliation{School of Physics and Technology, Wuhan University, Wuhan, Hubei 430072, China}

\author{Xiaoguang Wang}
\email{xgwang1208@zju.edu.cn}
\affiliation{Zhejiang Institute of Modern Physics and Department of Physics,
Zhejiang University, Hangzhou, Zhejiang 310027, China}
\affiliation{Graduate School of China Academy of Engineering Physics, Beijing, 100193, China}

\date{\today}% It is always \today, today,

\begin{abstract}
We propose to implement a solid-state rotation sensor by employing a many-body quantum spin system which takes the advantages of the easy controllability of the electron spin and the robustness provided by the collective nuclear spin state.
The sensor consists of a central electron spin coupled to many surrounding nuclear spins.
Previously, this central spin system has been suggested to realize a quantum memory.
Here, we further utilize the collective nuclear spins, which store a certain quantum state, to detect the macroscopic rotation.
Different from other nuclear spin-based gyroscopes, our proposal does not directly manipulate nuclear spins via nuclear magnetic resonance technique.
We analytically and numerically investigate the effects of partial nuclear polarization and decoherence on the sensitivity.
We also briefly introduce the procedure to generate entanglement between nuclear spins through the quantum memory technique and to utilize this entanglement to enhance the sensing performance.
Our proposal paves the way to the experimental realization of a compact solid-state, full-electrical and spin-based gyroscope.
%\begin{description}
%\pacs{03.65.Vf, 06.30.Gv, 61.72.jn}
%\end{description}
\end{abstract}

\maketitle

%\tableofcontents

\section{Introduction}
Quantum spin systems are attractive candidates to implement a quantum gyroscope, such as the well-explored nuclear magnetic resonance gyroscopes (NMRG)~\cite{donley2010nuclear,kitching2011atomic,kornack2005nuclear}, which have manifested excellent sensitivity in a laboratory.
However, these nuclear spin systems are usually gas based~\cite{kitching2011atomic,lam1983application} (the active sensing volume $>10\text{ mm}^3$), that are difficult to be miniaturized which limits their practical applications.
Recently, solid-state quantum spin systems, such as nitrogen-vacancy (NV) centers in a diamond, have been proposed to realize a rotation sensor via the geometric phase~\cite{maclaurin2012measurable,ledbetter2012gyroscopes,kowarsky2014non}.
These solid-state spin gyroscopes are promising to be miniaturized~\cite{ajoy2012stable} (the active sensing volume $<1\text{ mm}^3$) and to reduce the power consumption.
Moreover, the rotating solid-state quantum spin system has been investigated in recent experiments~\cite{wood2017magnetic,wood2018quantum}, and shown good prospect as a rotation sensor.

Unlike the spin-based magnetometry~\cite{budker2013optical}, the relative phase change in a Ramsey-like protocol is independent of the gyromagnetic ratio in a spin-based gyroscope.
Therefore, a nuclear spin is a better candidate than an electron spin as a rotation sensor, due to its stability to magnetic noise and long coherence time~\cite{budker2013optical,ledbetter2012gyroscopes}.
However, since the nuclear gyromagnetic ratio is much smaller than the electronic gyromagnetic ratio, $\gamma_n\ll \gamma_e$, as well as the nuclear spins are always well-isolated from the environment, the manipulation and measurement of nuclear spins are usually inefficient compared to electronic spins, which greatly limits the gyroscope sensitivity in practice~\cite{budker2013optical}.
An attractive way is to utilize the hyperfine interaction to manipulate and measure the nuclear spin state through the electron.
Single electronic spin coupled with single nuclear spin has been suggested to implement such a gyroscope in diamond~\cite{ajoy2012stable}.
However, the required pulse sequences of this protocol are rather complicated, consisting of both nuclear magnetic resonance (NMR) and electron spin resonance (ESR).
In addition, the single nuclear spin proximal to the NV electronic spin is also very sensitive to external perturbations~\cite{jiang2009repetitive}.

Usually, the nuclear spins inside the semiconductor quantum dot are factors that lead to the decoherence of electron spin~\cite{khaetskii2002electron,merkulov2002electron,coish2004hyperfine}, which is typically used as the qubit.
However, when being properly controlled, these nuclear spins can be useful resources for quantum computation or quantum sensing.
The long-lived quantum memory proposed by use of a semiconductor quantum dot, takes the advantages of the fast electron spin manipulation and the long coherence time provided by nuclei~\cite{taylor2003long,chekhovich2015suppression}.
Specifically, this protocol circumvents the difficulty of controlling single nuclear spins by utilizing collective nuclear spin state.
In this paper, we combine the quantum memory technique with the nuclear spin rotation sensing, to implement a quantum memory assisted rotation sensor.
The most significant advantage of our protocol is that, it only needs the fast and high-efficient electron spin manipulations and measurement instead of the slow and inefficient nuclear spin manipulations~\cite{slichter2013principles}.
Furthermore, we also show that, the performance of the rotation sensor can be enhanced by utilizing the nuclear spin entanglement, which is generated using the quantum memory technique as well.
\begin{figure*}
\includegraphics[width=0.8\textwidth]{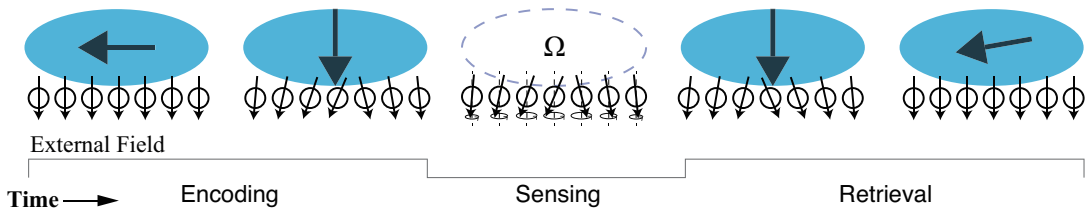}
\centering
\caption{\label{fig:qm_gyro}(color online) Schematic of the basic quantum memory assisted rotation sensing using a quantum dot.
At the start of the encoding (retrieval) stage, an electron spin polarized along the $x$ axis ($-z$ axis) is injected into the quantum dot, with an applied external magnetic field tuned on resonance.
During the sensing stage, the electron can be ejected and the nuclear spins experience a pseudo magnetic field due to a macroscopic rotation in the rotating frame.}
\end{figure*}

\section{Basic rotation sensing protocol}
We illustrate our proposal by use of the many-spin system in a semiconductor quantum dot, as depicted in Fig.~\ref{fig:qm_gyro}.
The protocol includes three stages, the encoding, the sensing and the retrieval stage.
At the beginning of the encoding stage, an electron in a certain spin state [here, the spin is polarized along $x$ axis, $\left|\psi_e(0)\right\rangle=(1/\sqrt{2})(\left|\uparrow\right\rangle_e+\left|\downarrow\right\rangle_e$)] is injected into a quantum dot.
With an external magnetic field properly tuned on resonance, the electron spin state is mapped onto the nuclear spins' collective state after a half period of Rabi oscillation~\cite{taylor2003long}.
This state mapping helps to build up the coherence of the collective nuclear spin state.
Then, the electron is ejected from the quantum dot.
The nuclear spins experience a pseudo magnetic field due to the macroscopic rotation and undergo a coherent precession in the rotating reference frame during the sensing stage.
After the rotation sensing, a new electron polarized along the $-z$ axis ($\left|\downarrow\right\rangle_e$) is injected, along with the external magnetic field tuned on resonance again.
After another half period of Rabi oscillation, the collective nuclear spin state with rotational information encoded, is mapped back into the electron spin and the electron spin state is measured subsequently.

We illustrate the protocol described above in much more detail using the \textit{central spin model}, which can describe a central electron spin coupled to many surrounding nuclear spins via hyperfine interaction in a semiconductor quantum dot.
For simplicity but without generality, we only consider nuclei with spin $I=1/2$ here.
During the short encoding and retrieval stages, the Hamiltonian is,
\begin{eqnarray}
\label{eq:qm}
H &=& g_e\mu_BB_0S_z+g_n\mu_nB_0\sum_{j=1}^NI_{jz}+\sum_{k=1}^N{A_kS_zI_{kz}}\notag\\
&+& \sum_{k=1}^N{\frac{A_k}{2}(S_{+}I_{k-}+S_-I_{k+})}.
\end{eqnarray}
The first term and the second term are the Zeeman energies of the electron spin and nuclear spins, respectively, where $g_e$ ($g_n$) is the Land\'{e} g-factor of the electron (nucleus), $\mu_B$ ($\mu_n$) is the Bohr magneton (nuclear magneton), and $B_0$ is the magnitude of the external magnetic field applied along $z$ axis.
The third (fourth) term is the dephase (flip-flop) term of the hyperfine contact interaction of the electron with $N$ surrounding nuclei, where $\mathbf{S}=(S_x,S_y,S_z)$ [$\mathbf{I}_k=(I_{kx},I_{ky},I_{kz})$] is the spin operator of the electron ($k$-th nucleus) with $S_{\pm}=S_x\pm iS_y$ and $I_{k\pm}=I_{kx}\pm iI_{ky}$.
Typically, the hyperfine coupling coefficient $A_k$ is nonuniform, for example, in a quantum dot, $A_k\propto |\psi(\mathbf{r}_k)|^2$, where $|\psi(\mathbf{r}_k)|^2$ is the electron profile density at site $\mathbf{r}_k$ of the $k$-th nucleus. 
This nonuniformity ($A_k\neq A$) intrinsically limits the fidelity of quantum memory~\cite{dobrovitski2006long} and will also limit the sensitivity of the rotation sensing proposed in this paper.
Specifically, for the encoding and retrieval processes to work, an external magnetic field needs to be applied to bring the system into resonance\footnote{The system can also be brought into resonance through electron g factors engineering~\cite{salis2001electrical,matveev2000g} or through optical ac stark shifts~\cite{salis2001origin}.} (flip-flop term of the hyperfine interaction dominates~\cite{taylor2003long}), $B_0=P\sum_k{A_k/(2g_e\mu_B)}-M_3/(2M_2g_e\mu_B)$, where $M_n=\sum_kA_k^n$ is defined as the $n$-th moment of the distribution of $A_k$, and $P$ is the average nuclear polarization.

During the relatively long rotation sensing stage, the evolution of the system is governed by,
\begin{equation}
\label{eq:Hs}
H_S=g_n\mu_n\mathbf{B_\Omega}\cdot\sum_{k=1}^N{\mathbf{I}_k}+\sum_{j<k}^N\Gamma_{jk}(3I_{jz}I_{kz}-\mathbf{I}_j\cdot\mathbf{I}_k).
\end{equation}
The first term describes nuclear spins precessing in the common pseudo magnetic field, $\mathbf{B_\Omega}=\mathbf{\Omega}/(g_n\mu_n)$, where $\mathbf{\Omega}$ is the rotational vector, in the frame rotating at an angular frequency $\Omega$.
For simplicity, we assume the rotational vector to be directed along the $z$ axis, so the first term becomes, $\Omega\sum_kI_{kz}$.
The second term describes the dipole-dipole interaction between nuclear spins, where the coefficient $\Gamma_{jk}$ is the dipolar coupling strength.
This term is neglected in $H$ (Eq.~\ref{eq:qm}), since the rate of state transfer ($\sim$GHz) is many orders of magnitude faster than the typical decoherence rate ($\sim$kHz) induced by the nuclear dipole-dipole interaction~\cite{taylor2003long}. 
However, during the rotation sensing stage, we can not neglect the nuclear dipole-dipole interaction any more, since for high sensitivity purpose, the sensing time usually needs to be as long as possible, which is ultimately restricted by the decoherence.

To demonstrate the entire rotation sensing protocol, as shown in Fig.~\ref{fig:qm_gyro}, we first consider the simplest case that nuclear spins are perfectly polarized, namely, $P=1$.
In this case, the dipole-dipole interaction between nuclear spins becomes greatly suppressed~\cite{fischer2009spin}, so we can temporarily neglect the nuclear dipole-dipole term, $H_S\approx \Omega\sum_{k=1}^NI_{kz}$.
Because of the conservation of the total spin, $[J_z,H]=[J_z,H_S]=0$, where $J_z=S_z+\sum_{k=1}^N{I_{kz}}$, the Hamiltonians can be expanded in a set of orthonormal basis states:
 $|\psi_0\rangle=\left|\downarrow\right\rangle_e\otimes\left|\phi_0\right\rangle$, $|\psi_1\rangle=\left|\uparrow\right\rangle_e\otimes\left|\phi_0\right\rangle$, $|\psi_2\rangle=(1/\sqrt{M_2})\sum_kA_k\left|\downarrow\right\rangle_e\otimes\left|\phi_k\right\rangle$, and $|\psi_3\rangle=|u^\prime\rangle/||u^\prime||$, where $|\phi_0\rangle=\left|\downarrow\downarrow\downarrow...\right\rangle_n$ and $|\phi_k\rangle=I_{k+}|\phi_0\rangle$ are collective nuclear spin states.
The state,
\begin{equation*}
|u^\prime\rangle=-\frac{1}{2\sqrt{M_2}}\sum_{k=1}^NA_k^2\left|\downarrow\right\rangle_e\otimes\left|\phi_k\right\rangle-\frac{M_3}{M_2}|\psi_2\rangle,
\end{equation*}
stands for the leakage from the ideal two-level system (when only considering the spin-exchange term during state transfer~\cite{taylor2003long}), and $||u^\prime||=\sqrt{M_4/M_2^2-M_3^2/M_2^3}$ is the norm of $|u^\prime\rangle$.
Thus the Hamiltonian $H$ can be represented as,
\begin{equation}
H=\frac{\sqrt{M_2}}{2}\begin{bmatrix}
\theta & 0 & 0 & 0\\
0 & 0 & 1 & 0\\
0 & 1 & 0 & \sqrt{2}\delta\\
0 & 0 & \sqrt{2}\delta & 0\\
\end{bmatrix}
\end{equation}
where $\delta=||u^\prime||/\sqrt{2}$ and $\theta=\sqrt{M_3^2/M_2^3}$.
In realistic physical systems, for example, in a semiconductor quantum dot, $A_k$ is often supposed in a Gaussian distribution, then $\delta, \theta \sim 1/\sqrt{N}$, which are small quantities when $N$ is large (the typical number of nuclear spins in a quantum dot is $\sim 10^4$--$10^6$).
The initial state of the compound system is $|\Psi(0)\rangle=\left|\psi_e(0)\right\rangle\otimes|\phi_0\rangle$.
After the encoding stage, the state of the total system evolves into,
\begin{equation}
\label{eq:state_after_encode}
|\Psi(t_E)\rangle=e^{-iHt_E}|\Psi(0)\rangle=\left| \uparrow\right\rangle_e\otimes\left|v_1\right\rangle_n+\left| \downarrow\right\rangle_e\otimes\left|v_2\right\rangle_n,
\end{equation}
where $\left|v_1\right\rangle_n$ and $\left|v_2\right\rangle_n$ denote the nuclear spin state.
Then the electron is ejected from the quantum dot (for example via tunneling into the adjacent electron reservoir in a gate-defined quantum dot~\cite{hanson2007spins}), which is equivalent to the Von Neumann's projection~\cite{dobrovitski2006long}.
After that, the system evolves into a mixed state,
\begin{equation}
\label{eq:state_after_ejection}
\rho(t_E)=\left|\uparrow\right\rangle\left\langle\uparrow\right|_e\otimes\left|v_1\right\rangle\left\langle v_1\right|_n+\left|\downarrow\right\rangle\left\langle\downarrow\right|_e\otimes\left|v_2\right\rangle\left\langle v_2\right|_n.
\end{equation}
During the sensing stage, the nuclear spins evolve under the Hamiltonian $H_S$,
\begin{eqnarray*}
\left|v_1^\prime\right\rangle_n=e^{-iH_St_S}\left|v_1\right\rangle_n,\\
\left|v_2^\prime\right\rangle_n=e^{-iH_St_S}\left|v_2\right\rangle_n.
\end{eqnarray*}
For the retrieval, another electron is injected and the state of the total system becomes,
\begin{equation}
\label{eq:state_after_injection}
\rho(t_E+t_S)=\left|\downarrow\right\rangle\left\langle\downarrow\right|_e\otimes(\left|v_1^\prime\right\rangle\left\langle v_1^\prime\right|_n+\left|v_2^\prime\right\rangle\left\langle v_2^\prime\right|_n).
\end{equation}
Finally, the state after the retrieval process becomes,
\begin{equation}
\rho_F\equiv\rho(t_E+t_S+t_R)=e^{-iHt_R}\rho(t_E+t_S)e^{iHt_R}.
\end{equation}

The encoding (retrieval) time $t_{E(R)}=\pi/\sqrt{M_2}$, and $t_S$ is the rotation sensing time.
Then the expectation values of the electron spin can be calculated, $\langle S_i\rangle=\text{Tr}[\tau_eS_i]$, with $i=x,y,z$, where $\tau_e=\text{Tr}_n\rho_{F}$ is the reduced density matrix of the electron spin.
Usually, the electron spin level population is measured in experiment and the corresponding signal (here the probability that $\left|\uparrow \right\rangle_e$ is populated after applying an electron spin $\pi/2$ pulse) is,
\begin{align}
\label{eq:perfect_polarization}
&S=\frac{1}{2}+\text{Tr}[e^{-i\frac{\pi}{2}S_y}\tau_ee^{i\frac{\pi}{2}S_y}S_z]\notag\\
&=\frac{1}{2}-\frac{\sin ^2\left(\frac{\pi}{4}   \lambda\right) \left[1+\left(2\lambda^2-1\right) \cos \left(\frac{\pi}{2}   \lambda\right)\right]}{\lambda^4}\cos(\pi\theta-\Omega t_S),
\end{align}
where $\lambda=\sqrt{1+\delta^2}$.
This analytical result demonstrates that the retrieved electron spin signal indeed oscillates at the rotational angular frequency $\Omega$.
The finite value of $\delta$ prevents complete transfer of spin state, just as in the quantum memory protocol.
Specifically, we note that the finite value of $\theta$ introduces an extra phase in the oscillation.
This characteristic may be used to detect properties of the nuclear ensemble, such as estimating the size of the nuclear ensemble.
The analytical result in Eq.~\ref{eq:perfect_polarization} is plotted in Fig.~\ref{fig:dipolar}(d), which shows great agreement with numerical simulations (in Sec.~\ref{numerical}).

Although the above procedure includes the electron ejection and injection, actually, just the same as in the quantum memory protocol~\cite{taylor2003long}, after mapping the quantum state into nuclear spins, the electron can either be ejected from the dot or stay in the dot with taking the system out of resonance, $g_e\mu_BB_0\ll P\sum_kA_k/2$.
When the system is off-resonant, the flip-flop term of the hyperfine interaction gets greatly suppressed, while the dephasing term, $H_D= S_z\sum_k{A_kI_{kz}}$, will lead to nuclear spin dephasing.
However, this dephasing can be refocused by flipping electron spin using spin echo-like technique via fast ESR pulses~\cite{taylor2003long}.
Besides, when the electron is present in the quantum dot, there also exists the electron-mediated nuclear dipole-dipole interaction which reduces the nuclear spin coherence time as well.
However, this second-order effect can be greatly suppressed~\cite{klauser2008nuclear,wust2016role} by increasing the electron Zeeman splitting (through applying a large external magnetic field or increasing the nuclear spin polarization) or by increasing the quantum dot size (large $N$).
Interestingly, it has also been proven~\cite{kurucz2009qubit} that the existence of the electron spin in the quantum dot can be helpful to protect the nuclear spin coherence by making an energy gap between the collective nuclear spin storage states with other states.
Specifically, spin diffusion due to nuclear dipole-dipole interaction can be suppressed by this mechanism.

\section{Estimation of the sensitivity}
However, in realistic situations, the nuclear polarization is far from perfect ($P<1$), which limits the state transfer fidelity and reduces the signal contrast.
At the same time, the nuclear dipole-dipole interaction induced decoherence also intrinsically limits the sensitivity.
In the following part, we will investigate the effects of these imperfections on the performance of the basic rotation sensing protocol. 
We quantify the performance of this rotation sensing protocol by calculating the quantum Fisher information (QFI), which is defined~\cite{helstrom1976quantum} as, $F_Q=\text{Tr}[\tau_e L^2]$, where $L$ is the symmetric logarithmic derivative (determined by $\partial \tau_e/\partial \Omega=(1/2)[\tau_e L+L\tau_e]$).
For the retrieved electron spin state here, the QFI has a simple expression~\cite{zhong2013fisher},
\begin{equation}
\label{eq:QFI}
F_Q = \left\{
\begin{array}{ll}
|\partial_\Omega\mathbf{v}|^2+\frac{(\mathbf{v}\cdot\partial_\Omega\mathbf{v})^2}{1-|\mathbf{v}|^2} & \text{if } |\mathbf{v}| < 1,\\
|\partial_\Omega\mathbf{v}|^2 & \text{if } |\mathbf{v}|  = 1.
\end{array} \right. 
\end{equation}
where $\mathbf{v}=(\langle S_x\rangle, \langle S_y\rangle,\langle S_z\rangle)$ is the Bloch vector representing the retrieved electron spin state.
Using the quantum Cram\'{e}r-Rao bound~\cite{CRbound}, we can estimate the sensitivity from the QFI, 
\begin{equation}
\delta \Omega=\frac{\sqrt{t_S+t_M}}{C\sqrt{F_Q}},
\end{equation}
where $C$ is the coefficient measuring the readout efficiency of the electron spin state; $t_M$ is the dead time which is required for initialization, transfer, and readout of the quantum state.

\subsection{Effects of partial nuclear polarization\label{appendixa}}
In order to analytically investigate the effects of average nuclear polarization on sensitivity, we consider the case that the electron spin has a uniform coupling with nuclear spins, namely, $A_k=A$.
Thus we can use the collective nuclear spin operator, $\mathbf{I}=\sum_{k=1}^N\mathbf{I}_k$, and the Hamiltonian during the encoding and retrieval stage can be simplified as,
\begin{equation}
\label{eq:HQ}
H_Q=(g_e\mu_BB_0+A I_z)S_z+\frac{A}{2}(S_+I_-+S_-I_+).
\end{equation}
For simplicity, we also neglect the nuclear dipole-dipole interaction during the sensing stage, with $H_S\approx\Omega I_z$.
Since $[\mathbf{I}^2,H_Q]=[\mathbf{I}^2,H_S]=0$, the magnitude of the collective nuclear spin angular momentum $I_0$ is the constant of motion during the entire rotation sensing protocol.
We will first calculate the spin dynamics with collective nuclear spin state $\left|I_0,M_0\right\rangle$, where $I_0(I_0+1)$ is the eigenvalue of $\mathbf{I}^2$ and $M_0$ is the eigenvalue of $I_z$, and then average over the distribution of $I_0$ and $M_0$ corresponding to the thermal nuclear spin state due to partial nuclear polarization.

The initial state of the system is $|\psi(0)\rangle=\frac{1}{\sqrt{2}}(\left|\uparrow\right\rangle_e+\left|\downarrow\right\rangle_e)\otimes|I_0,M_0\rangle$, and the Hamiltonian $H_Q$ can be expanded in the basis $\left|\uparrow\right\rangle_e\otimes|I_0,M_0\rangle$, $\left|\uparrow\right\rangle_e\otimes|I_0,M_0+1\rangle$, $\left|\downarrow\right\rangle_e\otimes|I_0,M_0\rangle$ and $\left|\downarrow\right\rangle_e\otimes|I_0,M_0-1\rangle$, as,
\begin{equation}
\label{eq:Hamiltonian}
H_Q=\begin{bmatrix}
p_1 & 0 & q_2 & 0\\
0 & p_1-\frac{A}{2} & 0 & q_1\\
q_2 & 0 & -p_1-\frac{A}{2} & 0\\
0 & q_1 & 0 & -p_1\\
\end{bmatrix}
\end{equation}
with,
\begin{eqnarray*}
q_1&=&\frac{A}{2}\sqrt{(I_0+ M_0)(I_0- M_0+1)},\\
q_2&=&\frac{A}{2}\sqrt{(I_0- M_0)(I_0+ M_0+1)},\\
p_1&=&\frac{A}{2}M_0+\frac{1}{2}g_e\mu_B B_0.\\
\end{eqnarray*}

After the encoding process, the compound system evolves into the state in the same form as Eq.~\ref{eq:state_after_encode}, except now with $\left|v_1\right\rangle_n=\alpha_1|I_0,M_0\rangle+\alpha_2|I_0,M_0-1\rangle$ and $\left|v_2\right\rangle_n=\alpha_3|I_0,M_0+1\rangle+\alpha_4|I_0,M_0\rangle$, where $\alpha_i$'s are appropriate coefficients.
After the ejection of the electron spin, the compound system evolves into a mixed state in the same form as Eq.~\ref{eq:state_after_ejection}.
During the sensing stage, the nuclear spins precess in the common pseudo magnetic field, and after the rotation sensing the state evolves into,
%\begin{widetext}
\begin{eqnarray*}
\left|v_1^\prime\right\rangle_n = \alpha_1 e^{-iM_0\Omega t_S}|I_0,M_0\rangle+\alpha_2 e^{-i(M_0-1)\Omega t_S}|I_0,M_0-1\rangle,\\
\left|v_2^\prime\right\rangle_n =\alpha_4 e^{-iM_0\Omega t_S}|I_0,M_0\rangle+\alpha_3 e^{-i(M_0+1)\Omega t_S}|I_0,M_0+1\rangle.
\end{eqnarray*}
%\end{widetext}
While after the injection of a new polarized electron spin, the state of the compound system again has the same form as Eq.~\ref{eq:state_after_injection}.
The evolution during the retrieval stage is unitary,
\begin{align}
\rho_F&=e^{-iHt_R}\rho(t_E+t_S)e^{iHt_R}\notag\\
&=|\psi_1\rangle\langle\psi_1|+|\psi_2\rangle\langle\psi_2|,
\end{align}
where $|\psi_1\rangle=e^{-iHt_R}\left|\downarrow\right\rangle_e\otimes\left|v_1^\prime\right\rangle_n$ and $|\psi_2\rangle=e^{-iHt_R}\left|\downarrow\right\rangle_e\otimes\left|v_2^\prime\right\rangle_n$.

With the final state obtained, the expectation value of the electron spin can be calculated, 
\begin{equation}
\langle S_i \rangle^\prime=\text{Tr}[\rho_FS_i]=\langle \psi_1|S_i|\psi_1\rangle+\langle \psi_2|S_i|\psi_2\rangle,
\end{equation}
with $i=x,y,z$.
The last step is the average over the distribution of $I_0$ and $M_0$,
\begin{equation}
\label{eq:sxR}
\langle S_i\rangle=\sum_{I_0=0}^{N/2}\sum_{M_0=-I_0}^{I_0}w(I_0,M_0)\langle S_i\rangle^\prime,
\end{equation}
where $w(I_0,M_0)$ is the probability distribution of $I_0$ and $M_0$. 
For the thermal state with average nuclear polarization $P$,
%\begin{widetext}
\begin{eqnarray*}
w(&&I_0, M_0)=\\
&& C_N^{N/2-I_0}(\frac{1+P}{2})^{N/2-M_0}(\frac{1-P}{2})^{N/2+M_0}\frac{2I_0+1}{N/2+I_0+1},
\end{eqnarray*}
%\end{widetext}
where $C_N^{N/2-I_0}$ is the binomial coefficient.
For the situation of large nuclear polarization ($P\sim 1$) and many nuclear spins ($N\to \infty$), the contribution of $I_0=-M_0$ dominates, and the distribution $w(I_0,M_0)$ can be approximated by,
\begin{equation}
w(M_0)\approx \frac{1}{\sqrt{2\pi}\sigma}e^{-\frac{(M_0-\bar{M})^2}{2\sigma^2}}
\end{equation}
with $\bar{M}=-\frac{N}{2}P$ and $\sigma^2=\frac{N}{4}(1-P^2)$.
Under this approximation, Eq.~\ref{eq:sxR} becomes,
\begin{widetext}
\begin{equation}
\langle S_x\rangle=\lim_{N\to \infty}\int^{N/2}_{-N/2}\frac{1}{\sqrt{2\pi}\sigma}e^{-\frac{(M_0-\bar{M})^2}{2\sigma^2}}\frac{2M_0\cos[\frac{(M_0-\bar{M}-1)\pi}{2\sqrt{N}}-\Omega]\sin^2[\frac{\sqrt{M_0+M_0(\bar{M}+1)+\bar{M}^2}\pi}{2\sqrt{N}}]}{M_0+M_0(\bar{M}+1)+\bar{M}^2} dM_0,
\end{equation}
\end{widetext}
while $\langle S_y\rangle$ and $\langle S_z\rangle$ can be calculated similarly.
After obtained these electron spin expectation values, we can calculate the quantum Fisher information using Eq.~\ref{eq:QFI} and the result is shown in Fig.~\ref{fig:gyroscope02}(a).

\subsection{Effects of nuclear decoherence}
\label{appendixb}
In order to get a qualitative understanding, we first consider a simple situation, namely, during the encoding and retrieval stages the central electron spin interacts strongly with the nearest nuclear spin (with polarization $P$), which is decohered by other surrounding nuclear spins during the rotation sensing stage.
The effect of the surrounding nuclear spins (bath spins) to the nearest nuclear spin can be approximated by a random field $B_N(t)$, which can be described by the Ornstein-Uhlenbek process~\cite{slava09}, with the correlation function,
\begin{equation}
\langle B_N(t)B_N(0)\rangle=b^2\exp(-Rt),
\end{equation}
where $b=\sqrt{\sum_{k=2}^N{\Gamma_{1k}}}$ (we assume the nearest nuclear spin to be the first nuclear spin), and $R$ is the correlation decay rate, which is determined by the internal dynamics of the bath spins.
The Hamiltonian that describes the decoherence process of the nearest nuclear spin is,
\begin{equation}
H_d=B_N(t)I_{1z},
\end{equation}
and the evolution of the nearest nuclear spin state can be calculated, using
\begin{equation}
i\frac{\partial \rho_n}{\partial t}=[H_d,\rho_n].
\end{equation} 
The solution of this decoherence process can be expressed as a phase damping channel~\cite{zhong2013fisher},
\begin{equation}
\rho_n=s\rho_{n0}+r(\rho_{n0,11}\left|\uparrow\right\rangle\left\langle \uparrow\right|_n+\rho_{n0,22}\left|\downarrow\right\rangle\left\langle \downarrow\right|_n)
\end{equation}
where $\rho_{n0}$ and $\rho_n$ are the density matrix of the nearest nuclear spin before and after the phase decoherence process; $s$ represents the phase decoherence process and $r=1-s$.
Using the cumulant expansion method~\cite{kubo1962generalized}, we get,
\begin{equation}
s=\langle e^{-i\int_{0}^tB_N(t^\prime)dt^\prime}\rangle= \exp[\frac{b^2}{R^2}(1-e^{-Rt}-Rt)].
\end{equation}
For a slow bath ($R\ll b$), we get $s\approx e^{-b^2t^2/2}$, which shows as a Gaussian decay. 
For clarity, we denote $s\approx e^{-(t/T_2^*)^2}$, with decoherence time $T_2^*=\sqrt{2}/b$.
Following again the entire rotation sensing procedure introduced above, the reduced density matrix of electron spin can be calculated,
\begin{equation}
\tau_e=a\left|\uparrow\right\rangle\left\langle \uparrow\right|_e+(1-a)\left|\downarrow\right\rangle\left\langle \downarrow\right|_e+b^*\left|\uparrow\right\rangle\left\langle \downarrow\right|_e+b\left|\downarrow\right\rangle\left\langle \uparrow\right|_e,
\end{equation}
where,
\begin{eqnarray*}
a&=&\frac{-8P\lambda^2+2(P^3-P^2+3P+5)\lambda}{(P^2-2P+5)^2},\\
b&=&-\frac{2ie^{s-i\lambda(t_S\Omega-P\frac{\pi}{2})}}{P^2-2P+5},
\end{eqnarray*}
with $\lambda=\sin ^2\left(\frac{1}{4} \pi  \sqrt{P^2-2P+5}\right)$.
Then the quantum Fisher information can be calculated as,
\begin{equation}
F_Q=\frac{16 t_S^2 e^{-\frac{2 t_S^2}{(T_2^*)^2}} \sin ^4\left(\frac{1}{4} \pi  \sqrt{P^2-2P+5}\right) }{(P^2-2P+5)^2}.
\end{equation}
From this result, we see that the decoherence of the nuclear spin seems to transfer to the electron spin and the performance of the rotation sensor is now limited by the nuclear decoherence time, instead of by the electron spin decoherence time.
Using the quantum Cram\'{e}r-Rao bound, we can estimate the sensitivity from the quantum Fisher information after a straightforward calculation,
\begin{equation}
\label{eq:phasedamping}
\delta \Omega=\frac{e^{(\frac{t_S}{T_2^*})^2}\sqrt{t_S+t_M}(P^2-2P+5)}{4Ct_S\sin ^2\left(\frac{\pi}{4} \sqrt{P^2-2P+5}\right)},
\end{equation}
where $T_2^*$ represents the characteristic nuclear decoherence time.
This equation indicates that there exists an optimal sensing time, $t_S\approx T_2^*/2$ (when $t_M\ll t_S$), to achieve the best sensitivity.

\subsection{Numerical investigation of the effects of partial polarization and decoherence}
\label{numerical}

In order to study these imperfections at the same time, we use exact numerical method, based on the Chebyshev polynomial expansion of the evolution operator~\cite{zhang2007modelling}, to simulate the dynamics of this many-spin system (using the full Hamiltonian in Eq.~\ref{eq:qm} and Eq.~\ref{eq:Hs} without any approximation).
Here, we follow the configurations used in previous work~\cite{dobrovitski2006long}, where $N=20$ nuclei are placed on a $4\times 5$ 2D lattice, with the value of $A_k$'s in the range of 0.31 to 0.96, corresponding to a Gaussian $|\psi(\mathbf{r})|^2$, and obtain $\Gamma_{jk}$'s from uniformly distributed random numbers in the range of -0.01 to 0.01.

\begin{figure}
\includegraphics[width=0.5\textwidth]{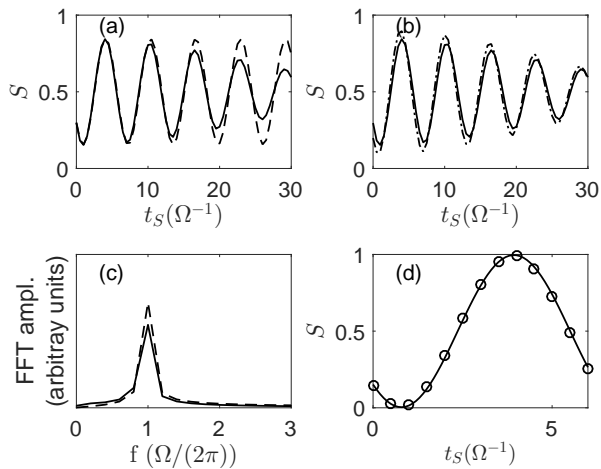}
%\includegraphics[width=0.8\textwidth]{dipolar.eps}
%\centering
\caption{\label{fig:dipolar}(a) Evolution of the electron spin population signal, calculated using exact numerical method with $N=20$ nuclear spins under homogeneous nuclear polarization with $P=0.8$. The solid (dashed) line corresponds to the case with (without) consideration of the nuclear dipole-dipole interaction. (b) Same as (a), except the dash-dotted line corresponds to the inhomogeneous nuclear polarization. (c) Fourier transform (FT) of the curves in (a). (d) The analytical result (solid line) and numerical result (circles) for the case of perfect nuclear polarization with $N=20$. The phase shift is due to the finite value of $\theta$.}
\end{figure}

Specifically, we will consider two nuclear polarization circumstances in detail.
Thermal nuclear polarization, which usually needs applying a large static magnetic field at low temperature~\cite{slichter2013principles}, results in homogeneous nuclear polarization and the initial nuclear state is $\rho_n(0)=(1/Z)\exp(-\gamma \sum_{k}{I_{kz}})$, where $Z$ is the partition function and $\gamma=2\tanh^{-1}(P)$.
On the other hand, dynamic nuclear polarization (DNP), for example, by passing a series of spin-polarized electrons through a quantum dot, leads to inhomogeneous nuclear polarization~\cite{petta2008dynamic,wu2016inhomogeneous} and the initial nuclear state is $\rho_n(0)=\otimes_{k=1}^N\rho_{nk}(0)$, with $\rho_{nk}(0)=(1/2)\mathbf{1}+p_kI_{kz}$, where $p_k=1-\exp(-2\beta A_k^2)$ is the polarization of the $k$-th nuclear spin ($\beta$ is a parameter reflecting an effective spin temperature) and the average nuclear polarization is defined as, $P=(1/N)\sum_{k=1}^Np_k$.
The evolution of the entire rotation sensing process can be calculated following the procedures introduced above and here the initial state of the compound system is $\rho(0)=\left|\psi_e(0)\right\rangle\left\langle\psi_e(0)\right|\otimes\rho_n(0)$.

The numerically calculated electron spin population $S$ versus sensing time $t_S$ is plotted in Fig.~\ref{fig:dipolar}(a) and (b).
Indeed, the nuclear dipole-dipole interaction results in decay of the signal, restricting the available sensing time and eventually limiting the sensitivity.
In Fig.~\ref{fig:dipolar}(c), we plot the Fourier analysis of the signals in Fig.~\ref{fig:dipolar}(a).
It shows that, the peak frequency, which corresponds to the rotational frequency, is broadened but not shifted by the nuclear dipole-dipole interaction.

\begin{figure}
\includegraphics[width=0.5\textwidth]{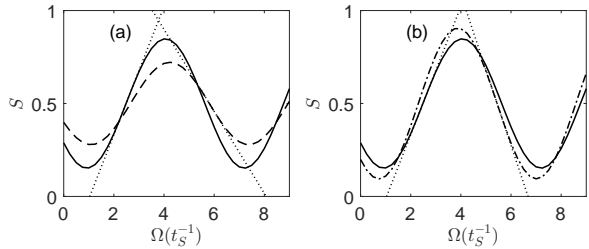}
\centering
\caption{\label{fig:gyroscope}Numerical results of the electron spin population signal versus the rotational angular frequency with a fixed sensing time $t_S< T_2^*$, for different nuclear polarizations in a slope detection measurement. The dotted lines reflect the sensitivity. (a) Comparison of different degrees of nuclear polarization, $P=0.8$ (solid line) and $P=0.6$ (dashed line) for the homogeneous case. (b) Comparison of homogeneous nuclear polarization (solid line) and inhomogeneous nuclear polarization (dash-dotted line) with the same $P=0.8$.}
\end{figure}

Next, we plot the electron spin population $S$ as a function of the rotational angular frequency $\Omega$ with a fixed sensing time $t_S$ for different nuclear polarizations in Fig.~\ref{fig:gyroscope}(a) and (b).
The high degree of nuclear polarization leads to a large oscillation amplitude, as the result shown in Fig.~\ref{fig:gyroscope}(a).
Meanwhile, as shown in Fig.~\ref{fig:gyroscope}(b), with the same average nuclear polarization, the inhomogeneous case presents larger oscillation amplitude than the homogeneous case, indicating higher signal-to-noise ratio in experiment.
These features are similar to that of the quantum memory protocol~\cite{ding2014high}, since the increase of the state transfer fidelity will naturally lead to an improvement of the signal contrast and the sensitivity.

\begin{figure}
\includegraphics[width=0.5\textwidth]{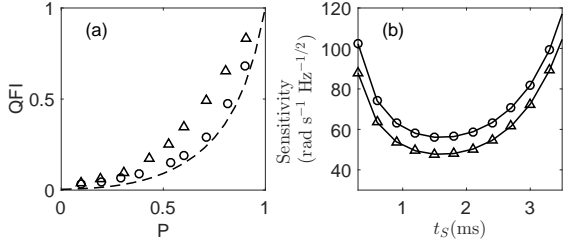}
\centering
\caption{\label{fig:gyroscope02}(a) Quantum Fisher information as a function of average nuclear polarization with a fixed sensing time. The circles (triangles) are numerical results for the homogeneous (inhomogeneous) nuclear polarization. The dashed line corresponds to the analytical result of the uniformly coupling model ($A_k=A$, see Sec.~\ref{appendixa}). (b) The estimated sensitivity for the inhomogeneous case (triangles) and the homogeneous case (circles) corresponding to the data in Fig.~\ref{fig:dipolar}(b) by assuming $C=0.8$, $T_2^*= 3\text{ ms}$, and $t_M= 100\text{ ns}$.}
\end{figure}

The QFI as a function of average nuclear polarization with fixed sensing time is depicted in Fig.~\ref{fig:gyroscope02}(a).
Apparently, the QFI increases as the average nuclear polarization increases, indicating an improvement in measurement precision.
In addition, under the same average nuclear polarization, the inhomogeneous case exhibits much larger QFI than the homogeneous case, almost 2 times larger when $P=0.7$.
We also find that the behavior of the homogeneous case is in accordance with the analytical model (by assuming $A_k=A$, see Sec.~\ref{appendixa}), whereas the inhomogeneous case shows nontrivial characteristics, which can be explained by the utilization of the nonuniformity of the hyperfine coupling~\cite{ding2014high}.

Finally, we investigate the effects of decoherence on the sensitivity.
In Fig.~\ref{fig:gyroscope02}(b), we present the estimated sensitivity from the exact numerical simulations.
As we can see, there indeed exists an optimal sensing time, $t_S\approx T_2^*/2$, where $T_2^*$ can be obtained by fitting to a Gaussian decay (see Sec.~\ref{appendixb}) of the corresponding signal in Fig.~\ref{fig:dipolar}(b).
Again, the inhomogeneous case always exhibits an enhanced sensitivity compared with the homogeneous case.
In other words, the inhomogeneous polarization method relaxes the requirement on average nuclear polarization to obtain the same sensitivity as using the homogeneous polarization method.
Besides, to obtain the inhomogeneous nuclear polarization through DNP usually does not make the stringent requirement on temperature and magnetic fields as the conventional thermal polarization method does.

\subsection{Feasibility of the basic protocol}
Our proposal can be implemented using quantum dots of various materials and sizes.
%There are two major varieties of semiconductor quantum dots, namely, the gate-defined quantum dot and the self-assembled quantum dot.
%The gate-defined quantum dot are usually controlled electrically at very low temperatures ($\sim 100$ mK), while the self-assembled quantum dots are usually controlled optically at higher temperatures ($\sim 4$ K).
The maximum average nuclear polarization achieved so far is $\sim 80\%$~\cite{chekhovich2017measurement} in a nanohole in-filled GaAs/AlGaAs quantum dot using optical pumping method with an external magnetic field $\sim 8 \text{ T}$, operating at temperature $\sim 4.2 \text{ K}$.
The size of this in-filled GaAs/AlGaAs quantum dot is $\sim 50\text{ nm}$, with $\sim 10^5$ nuclei inside the quantum dot and nuclear spin relaxation time $T_1>500 \text{ s}$ is reported in this experiment.
While in a gate-defined quantum dot~\cite{petersen2013large}, the nuclear spin polarization reaches $\sim 50\%$, using the electrical control, at operating temperature $\sim 100 \text{ mK}$.
In this experiment, the size of the GaAs double quantum dot is $\sim 100\text{ nm}$ and a single domain nanomagnet is used to apply the static magnetic field.

The dead time $t_M$ includes the time needed for the electron spin initialization and readout, the time needed for nuclear spin polarization and the time needed for quantum state transfer.
For the basic rotation sensing protocol introduced here, after one sensing cycle, only the electron spin state is measured and ideally the nuclear spins will return to the initial state.
Hence, in each sensing cycle, only the electron polarization is consumed and due to their extremely long relaxation time in a quantum dot~\cite{urbaszek2013nuclear}, the nuclear spins only need to be repolarized after many sensing cycles (the time needed for initial nuclear spin polarization is equivalent to the startup time of the apparatus).
Therefore, the dominate contribution to $t_M$ is the time needed for the electron spin initialization and readout, which can be as short as $\sim 100\text{ ns}$ ~\cite{barthel2010fast}, while the time needed for state transfer is $\sim 1\text{ ns}$~\cite{taylor2003long}.
Finally, with the electron spin readout visibility $\sim 80\%$~\cite{hanson2005single} and the nuclear decoherence time $T_2^*\sim 3$ ms~\cite{taylor2003long,mehring2012high,chekhovich2015suppression}, we can estimate the sensitivity as $\delta \Omega \sim 50\text{ rad s}^{-1}\text{ Hz}^{-1/2}$ per sensor unit (or per qubit), which is at the same level as the proposed solid-state rotation sensor using nuclear spins in diamond~\citep{ledbetter2012gyroscopes,ajoy2012stable}.

In order to obtain high sensitivity in practice, the basic protocol proposed here should make use of large wafers of quantum dots which confine spins as ensemble sensor.
One advantage of semiconductor quantum dot qubit is the scalability compared to many other qubit candidates and the manufacturing technology of semiconductors can be directly employed to fabricate quantum dots, for example, the gate-defined quantum dot can be lithographically defined by metallic gates on the semiconductor substrate.
In nowadays semiconductor industry, billions of transistors can be integrated on a chip with size $\sim 100 \text{ mm}^2$.
The projected sensitivity using QD ensemble with similar integration level on such a chip is $\sim 10^{-4}\text{ rad s}^{-1}\text{ Hz}^{-1/2}$ and here we suppose the sensitivity scales as $1/\sqrt{M}$ for the ensemble sensor, where $M$ is the number of quantum dots integrated on the chip.
Besides, the scaling problem for quantum sensing may be much easier to solve than for quantum computation, since qubits may need not be coupled together for quantum sensing purpose, while it is usually essential to couple qubits for large-scale quantum computation.

Some theories predicted that after the preparation by electron spins, the collective nuclear spin state may evolve into the dark state $\left|\mathcal{D}\right\rangle$~\cite{imamoglu2003optical,taylor03controlling}, which is defined by $\sum_k^NA_kI_{k-}\left|\mathcal{D}\right\rangle=0$.
The dark state has low values of polarization and purity, but it can still be used as long-lived quantum memory due to its symmetry property.
Our rotation sensing scheme can be generalized to this quantum memory protocol as well and it is estimated that the quantum state transfer fidelity can exceed $80\%$ with state transfer time $\sim 100\text{ ns}$ in a quantum dot containing $\sim 10^4$ nuclei, while the average nuclear polarization is vanishingly small~\cite{taylor03controlling}.
Moreover, the recently proposed collective nuclear spin quantum memory using a strain-enabled quantum dot~\cite{denning2019collective,gangloff2019quantum} predicted a state storage fidelity up to $90\%$ with nuclear polarization $\sim 50\%$.
Even though in this proposal the interaction used to encode and retrieve is different, the rotation sensing protocol proposed here can still be easily generalized.

\section{Entanglement enhanced rotation sensing}
In order to obtain the full quantum enhancement for the rotation sensing protocol proposed here, the entanglement between nuclear spins can be utilized~\cite{rudner2011generating,jones2009magnetic,cooper2008environment}.
The conventional method to generate entanglement between surrounding nuclear spins in a central spin system is firstly to employ the Hadamard gate on the central electron spin and then to employ the controlled-NOT gate, where the rotation of the surrounding nuclear spins depends on the central electron spin state~\cite{jones2009magnetic,schaffry2010quantum,goldstein2011environment,cappellaro2012environment}.
However, this technique is restricted by the electron spin coherence time, which is relatively short in semiconductor quantum dots.
Besides, the rotation of nuclear spins is driven by NMR, which is against the advantage of our scheme (only electron spin is directly manipulated in our scheme).
Instead, here we adopt the idea (mentioned in Ref.~\cite{taylor2003long}) based on quantum memory technique and employ the tailored procedure introduced in Ref.~\cite{law96arbitrary} to generate nuclear spin entanglement, which still only needs the fast and efficient electronic manipulations.
The basic idea lies in engineering the collective nuclear spin state through continuously mapping specific electron spin states onto nuclear spins in many cycles.
In each cycle, firstly the electron spin is prepared into a specific quantum state via ESR manipulations (the \textit{classical channel}), then the hyperfine interaction is tuned on-resonant and the electron spin state is encoded onto the collective nuclear spin state (the \textit{quantum channel}). 

We illustrate the basic entanglement generation procedure by considering the ideal case of uniform coupling ($A_k=A$) and perfect nuclear polarization ($P=1$).
Again, the collective nuclear spin state can be represented in the basis of the total nuclear spin operator, $\mathbf{I}=\sum_k{\mathbf{I}_k}$, which is the same as in Sec.~\ref{appendixa}, and the Hamiltonian describing the quantum channel is also $H_Q$ in Eq.~\ref{eq:HQ}.
By tuning the external magnetic field, $g_e\mu_BB_0\sim -A\langle I_z\rangle$, the system can be brought into resonance, with the flip-flop term dominating the dynamics.
While for the classical channel, the system is tuned off-resonant by changing the external magnetic field, with the flip-flop term being greatly suppressed and the Hamiltonian describing the classical channel becomes,
\begin{eqnarray}
H_C&&\cong(g_e\mu_BB_0+A I_z)S_z\notag\\
&&+g_e\mu_BB_1(e^{-i(\omega_1t+\phi_1)}S_++e^{i(\omega_1t+\phi_1)}S_-),
\end{eqnarray}
where $B_1$, $\omega_1$ and $\phi_1$ are the amplitude, frequency and phase of the applied alternating magnetic field, respectively.
In fact, using our entanglement generation technique, arbitrary collective nuclear spin state $|\Psi\rangle_n=\sum_{l=0}^{2I_0}{c_l|I_0,-I_0+l\rangle}$ can be prepared from the initially polarized state $|\Psi_0\rangle_n=|I_0=N/2,M_0=-I_0\rangle$.
However, for the rotation sensing purpose, here we only consider the preparation and utilization of the Greenberger-Horne-Zeilinger (GHZ)-like entangled state,
\begin{equation*}
|\Psi^*\rangle_n=\frac{1}{\sqrt{2}}(|I_0,M_0=-I_0\rangle+|I_0,M_0=-I_0+m\rangle),
\end{equation*}
and here $m\le N/2$ denotes the largest excitation in the entangled state.
Next, we will give a detailed description of the preparation of this entangled state.

It is much easier to consider the inverse state evolution of the entanglement generation, namely,
\begin{equation*}
\left|\downarrow\right\rangle_e\otimes|\Psi_0\rangle_n=C_1^\dagger Q_1^\dagger C_2^\dagger Q_2^\dagger\cdots C_m^\dagger Q_m^\dagger (\left|\downarrow\right\rangle_e\otimes|\Psi^*\rangle_n),
\end{equation*}
where the evolution operator $C_j$ in the $j$--th cycle is governed by $H_C$ and $Q_j$ is governed by $H_Q$.
For each cycle, these following equations should be satisfied,
\begin{eqnarray*}
(\langle I_0,-I_0+m-j|\otimes{_e\langle}\downarrow|)Q_{m-j}^\dagger |F_{j}\rangle=0, \\
(\langle I_0,-I_0+m-j-1|\otimes{_e\langle}\uparrow|)C_{m-j}^\dagger Q_{m-j}^\dagger |F_{j}\rangle=0, 
\end{eqnarray*}
where $1\leq j\leq m$ and,
\begin{equation*}
|F_{j}\rangle\equiv C_{m-j+1}^\dagger Q_{m-j+1}^\dagger C_{m-j+2}^\dagger Q_{m-j+2}^\dagger\cdots C_m^\dagger Q_m^\dagger |F_{0}\rangle,
\end{equation*}
with $|F_{0}\rangle\equiv \left|\downarrow\right\rangle_e\otimes |\Psi^*\rangle_n$.

As an example, we consider the first cycle of the evolution.
For the quantum channel, namely,
\begin{eqnarray*}
Q_m^\dagger\frac{1}{\sqrt{2}}\left|\downarrow\right\rangle_e\otimes(|I_0,-I_0\rangle+|I_0,-I_0+m\rangle)\rightarrow \\
\frac{1}{\sqrt{2}}(e^{-i\varphi_m}\left|\downarrow\right\rangle_e\otimes|I_0,-I_0\rangle+i\left|\uparrow\right\rangle_e\otimes|I_0,-I_0+m-1\rangle),
\end{eqnarray*}
we have to tune the external magnetic field in $H_Q$ with $g_e\mu_BB_0\approx -A(-I_0+m)$, so the flip-flop term becomes effective only for the $\left|\downarrow\right\rangle_e\otimes|I_0,-I_0+m\rangle$ subspace.
Meanwhile, for the $\left|\downarrow\right\rangle_e\otimes|I_0,-I_0\rangle$ subspace, it will evolve under $(g_e\mu_BB_0+A I_z)S_z$, resulting in an extra relative phase $\varphi_m=\frac{A}{2}mt_E$, with the encoding time $t_E=\pi/[A\sqrt{(m+1)(2I_0-m)}]$.
However, this extra phase can be eliminated in the last cycle (we will illustrate this later).
While for the classical channel, namely,
\begin{eqnarray*}
C_m^\dagger Q_m^\dagger\frac{1}{\sqrt{2}}\left|\downarrow\right\rangle_e\otimes(|I_0,-I_0\rangle+|I_0,-I_0+m\rangle)\rightarrow \\
\frac{1}{\sqrt{2}}(e^{-i\varphi_m}\left|\downarrow\right\rangle_e\otimes|I_0,-I_0\rangle+i\left|\downarrow\right\rangle_e\otimes|I_0,-I_0+m-1\rangle),
\end{eqnarray*}
we need to set appropriate parameters of the alternating magnetic field to selectively rotate the electron spin, for example, in this step, we need to set $\omega_1=A(I_0-m+1)/(g_e\mu_B\hbar)$, $\phi_1=0$ and $B_1=\pi/(g_e\mu_Bt_C)$, where $t_C$ is the duration of the ESR pulse.

The subsequent cycles are operated similarly, until the classical channel of the last cycle,
\begin{eqnarray*}
C_1^\dagger \frac{1}{\sqrt{2}}(e^{-i\sum_j\varphi_j}\left|\downarrow\right\rangle_e\otimes|I_0,-I_0\rangle+i^m\left|\uparrow\right\rangle_e\otimes|I_0,-I_0\rangle))\rightarrow \\
\left|\downarrow\right\rangle_e\otimes|I_0,-I_0\rangle,
\end{eqnarray*}
where we need to choose the correct phase $\phi_1$ in $C_1^\dagger$ to rotate the electron spin state $\frac{1}{\sqrt{2}}(e^{-i\sum_j\varphi_j}\left|\downarrow\right\rangle_e+i^m\left|\uparrow\right\rangle_e)$ to the ground state $\left|\downarrow\right\rangle_e$.

The preparation of the entangled state from the initial polarized state is the inverse process of the above evolution.
Compared to the conventional entanglement preparation method, here we only utilize the fast and efficient electron spin manipulation, which still keeps the advantage of our rotation sensing protocol.
However, the performance of this entanglement generation procedure hinges on the quality of ESR manipulation, especially for those very early cycles.

After preparation, the entangled collective nuclear spin state can be employed to sense rotation under the Hamiltonian, $H_S\approx\Omega I_z$, and the collective nuclear spin state evolves into,
\begin{equation}
|\Psi^*\rangle_n^\prime=\frac{1}{\sqrt{2}}(|I_0,-I_0\rangle+e^{im\Omega t_S}|I_0,-I_0+m\rangle).
\end{equation}
Using the inverse entangling process, the relative phase in the collective nuclear spin state which encoded the rotational information can be mapped back to the electron spin,
\begin{eqnarray*}
\left|\downarrow\right\rangle_e\otimes\frac{1}{\sqrt{2}}(|I_0,-I_0\rangle+e^{im\Omega t_S}|I_0,-I_0+m\rangle)\\
\rightarrow \frac{1}{\sqrt{2}}(\left|\uparrow\right\rangle_e+e^{im\Omega t_S}\left|\downarrow\right\rangle_e)\otimes|I_0,-I_0\rangle.
\end{eqnarray*}
Clearly, the accumulated relative phase using this entanglement enhanced protocol is $m$ times larger than that of the basic rotation sensing protocol without entanglement.
Besides, the nuclear spins return to the initial polarized state and another full sensing cycle can be repeated.
It indicates that, ideally, the sensitivity scales as $1/N$ (since $m\le N/2$) in a single quantum dot for the entanglement enhanced protocol, approaching Heisenberg limit, where $N$ is the number of nuclear spins in a quantum dot ($N\sim 10^4$--$10^6$ in typical quantum dots).
While for the entanglement enhanced ensemble sensor, the overall sensitivity scales as $1/(N\sqrt{M})$, where $M$ is the number of quantum dots.

However, much more factors need to be considered to estimate the overall sensitivity enhancement in practical situations.
Extra dead time to prepare and readout the entangled state should be included to estimate the sensitivity and this may lead to an optimal value of $m$ to prepare the entangled state. 
Usually, the prepared GHZ-like entangled state has a relatively short lifetime~\cite{krojanski2004scaling}, while other entangled states which are robust to decoherence may be prepared by modifying procedures introduced above~\cite{rudner2011generating}.
For realistic situations of nonuniform coupling strength ($A_k\neq A$) and partial nuclear polarization ($P<1$), the perfect GHZ-like entangled state is difficult to generate.
However, through similar procedures as introduced above, partially entangled collective nuclear spin state can still be prepared and used for quantum enhanced rotation sensing~\cite{goldstein2011environment,cappellaro2012environment}.

\section{Discussion}
In this paper, we propose a rotation sensing protocol utilizing the solid-state central spin system, which can be implemented using a semiconductor quantum dot.
Gas-based gyroscopes using various quantum systems, such as cold atoms~\cite{stockton2011absolute}, atom beam~\cite{gustavson1997precision}, NMRG~\cite{kornack2005nuclear}, and ring lasers~\cite{stedman2003detectability}, show excellent sensitivity, but usually require a large active sensing volume and high power consumption.
For example, the miniaturization of the vapor cell of NMRG will increase the cell-wall effects, resulting in reduction of the nuclear spin lifetime~\cite{budker2013optical}.
On the other hand, solid-state spin gyroscopes possess the advantage of miniaturization.
Specifically, miniaturization is important for magnetic shielding, which is crucial for the performance of spin-based sensors~\cite{budker2013optical}.

Our protocol also shows some advantages compared to other solid-state rotation sensing protocols.
%The fiber optic gyroscope also needs a large active sensing volume.
For instance, the ubiquitous microelectromechanical systems (MEMS) gyroscope suffers from the problem of sensitivity drifts due to the formation of charge asperities~\cite{jekeli2012inertial}, which does not occur in spin-based protocols.
In particular, compared to the recently proposed another solid-state spin-based protocol using NV centers in diamond~\citep{ajoy2012stable,ledbetter2012gyroscopes}, which own the advantage of room-temperature operation, the gyroscope using QD also shows some advantages as follows.
First, in the diamond, the single proximal $^{14}\text{N}$ nuclear spin has relatively short coherence time because of its strong interaction with the NV electronic spin (such as the optical illumination on NV centers will lead to nuclear spin depolarization~\cite{jiang2009repetitive,knowles2017controlling}), and the nuclear spin needs to be repolarized in every sensing cycle.
On the contrary, the electron can be removed from the quantum dot to switch off the hyperfine interaction completely, resulting in a remarkably long nuclear coherence time~\cite{urbaszek2013nuclear} and the nuclear spins only need to be repolarized after many sensing cycles due to the utilization of collective state and the long nuclear relaxation time in a quantum dot.
Second, our proposal circumvents direct nuclear manipulations via NMR technique by fully exploiting the coherent spin state transfer instead.
It is widely known that the nuclear spin manipulation (especially for single nuclear spins) is usually difficult and inefficient, as well as introducing extra noise to the system~\cite{slichter2013principles}.
Third, since there are much more nuclear spins interacting strongly with the electron spin in the quantum dot, the scheme proposed here seems to be more promising to employ quantum many-body entanglement for enhanced sensing.

By use of a gate-defined semiconductor quantum dot, a full-electrical spin-based rotation sensor seems feasible, since nuclear spins can be electrically polarized by DNP~\cite{reilly2008suppressing,petta2008dynamic}, while the electron spin can be coherently controlled by electric fields~\cite{kato2003gigahertz,nowack2007coherent,pioro2008electrically} and readout via spin-to-charge conversion~\cite{elzerman2004single}.
Because electrical fields can be easily confined in nanoscale regions compared to magnetic fields or optical fields, qubits in the full-electrical solutions can be locally addressed and controlled, which is beneficial for the generation of quantum entanglement and important for large scale quantum devices.
Besides, full-electrical solutions can be much more compact and easier to integrate with other quantum devices than optical solutions.

In principle, the technique proposed in this paper may be generalized to various quantum systems that can be implemented as quantum memory, such as atomic ensembles~\cite{kozhekin2000quantum,julsgaard2004experimental,zhao2009long}, rare-earth-ion doped crystals~\cite{hedges2010efficient}, cold ions~\cite{kielpinski2001decoherence,langer2005long}, or other solid-state spin systems~\cite{morton2008solid,dutt2007quantum,steger2012quantum,maurer2012room}.
Particularly, it has been proposed to store the quantum state of light into nuclear spins inside a quantum dot~\cite{schwager2010quantum}.
Our rotation sensing scheme can be directly adapted to this quantum interface, with the quantum state mapping between light and nuclear spins, while still employing nuclear spins as the rotation sensor.
%Higher quantum state storage fidelity and longer storage time will naturally lead to a better sensitivity.
In addition to the utilization of entanglement between nuclear spins, the electron spins in different quantum dots can also be entangled together via the exchange interaction to further enhance the rotation sensing~\cite{goldstein2011environment,cappellaro2012environment}. 

%\textit{Physical Review} style requires that the initial citation of
%figures or tables be in numerical order in text, so don't cite
%Fig.~\ref{fig:wide} until Fig.~\ref{fig:epsart} has been cited.
%
\begin{acknowledgments}
This work was supported by the National Key Research and Development Program of China (Grants No. 2017YFA0304202 and No. 2017YFA0205700), the NSFC through Grant No. 11875231 and No. 91836101, and the Fundamental Research Funds for the Central Universities through Grant No. 2018FZA3005.
\end{acknowledgments}

\appendix

%\nocite{*}
\bibliographystyle{apsrev4-1}

\end{document}